\documentclass{IEEEtran}
\usepackage{siunitx}
\usepackage{commath}
\usepackage{graphicx}
\usepackage{textcomp}
\usepackage{cite}
\usepackage{subcaption}

\begin{document}
%\bstctlcite{IEEEexample:BSTcontrol} 
\title{Photonic convolutional neural networks using integrated diffractive optics}

\author{\IEEEauthorblockN{Jun Rong Ong, 
		Chin Chun Ooi, 
		Thomas Y. L. Ang, 
		Soon Thor Lim, 
		Ching Eng Png}

\thanks{%Manuscript received Month D, YYYY; revised Month DD, YYYY.
%J.R. Ong, T.Y.L. Ang, S.T. Lim and C.E.Png are supported by A*STAR-NTU-SUTD AI Partnership Grant (RGANS1901). 
J.R. Ong, C.C. Ooi, T.Y.L. Ang, S.T. Lim and C.E. Png are with the Institute of High Performance Computing, Agency for Science, Technology and Research (A*STAR), 138632, Singapore. (Corresponding author: J. R. Ong, e-mail: ongjr@ihpc.a-star.edu.sg)
}}

\IEEEtitleabstractindextext{
\begin{abstract}
 With recent rapid advances in photonic integrated circuits, it has been demonstrated that programmable photonic chips can be used to implement artificial neural networks. Convolutional neural networks (CNN) are a class of deep learning methods that have been highly successful in applications such as image classification and speech processing. We present an architecture to implement a photonic CNN using the Fourier transform property of integrated star couplers. We show, in computer simulation, high accuracy image classification using the MNIST dataset. We also model component imperfections in photonic CNN and show that the performance degradation can be recovered in a programmable chip. Our proposed architecture provides a large reduction in physical footprint compared to current implementations as it utilizes the natural advantages of optics and hence offers a scalable pathway towards integrated photonic deep learning processors. 
\end{abstract}

\begin{IEEEkeywords}
% List of suggested keywords: http://www.ieee.org/organizations/pubs/ani_prod/keywrd98.txt
Artificial neural networks, Neuromorphics, Photonic integrated circuits, Silicon photonics
\end{IEEEkeywords}}

\maketitle
\IEEEdisplaynontitleabstractindextext

\section{Introduction}

Deep learning methods like CNNs have received a huge amount of interest from the research community as well as the general public after it was shown to approach human level performance in image recognition tasks \cite{lecun2015deep}. This breakthrough was due in part to the availibility of fast graphical processing units (GPUs) that greatly accelerated the implementation of deep neural networks \cite{krizhevsky2012imagenet}. Recent efforts in developing hardware machine learning accelerators include massively parallel high-throughput devices \cite{jouppi2018motivation}, as well as neuromorphic computing architectures in which aspects of the design mimic principles present in biological neural networks \cite{furber2014spinnaker,essera2016convolutional}. The search for alternative computing paradigms is also fueled by the impending end of Moore's law, which is a result of the fundamental limits of transistor scaling, as well as related bottlenecks in power dissipation and interconnect bandwidth \cite{cavin2012science}.    

On the other hand, optical computing platforms potentially offer a number of attractive advantages such as parallelism through wavelength and temporal multiplexing \cite{tait2017neuromorphic,brunner2013parallel} as well as ideally non-dissipative interconnects \cite{beausoleil2008nanophotonic}. Such potential advantages could be harnessed in high-performance computing systems, such as dedicated hardware accelerators for machine learning. Artificial neural networks are particularly suited for optical implementation as they mainly rely on computing matrix multiplications, which can be performed with high speed and throughput with optics \cite{goodman1978fully,yang2012chip}. In fact, the inherent advantages of optical neural networks were studied in detail decades ago using bulk optics \cite{caulfield1989optical}. Recent progress in photonic integrated circuits and programmable photonics has enabled the demonstration of a chip-scale optical neural network (ONN) using a mesh of interferometers and phase shifters \cite{shen2017deep}. The authors used the abilty of such meshes to implement general unitary transformations, and together with a singular value decomposition \cite{miller2013self}, showed a ONN equivalent to a fully-connected multi-layer perceptron (MLPs). CNNs, in contrast to such fully-connected networks, take advantage of hierarchical patterns in the underlying data by having shared weights between network nodes (i.e. a convolution operation) and hence have a reduced scale of complexity and connectedness. The CNN architecture takes its inspiration from the animal visual cortex \cite{lecun2015deep} and is in that sense a neuromorphic computing system. 

Coherent optical information processing systems, for example optical correlators, rely heavily on the natural Fourier transform property of optics \cite{goodman2005introduction}. Conventionally, the correlation filters are hand designed by human experts. More recently, diffractive optics systems have demonstrated a high degree of success in image classification tasks \cite{lin2018all,yan2019fourier}. Such systems are constructed using sequential amplitude and phase masks, with each individual pixel in the masks being a trainable parameter in a deep learning optimization algorithm. In this way, the filter masks are generated automatically using machine learning. With the lenses arranged in a ``4f" system, the optics performs convolutions in a similar fashion as in the convolution layers in a CNN. However, most proposals of optical CNNs have focused on a hybrid optical-electronic system configuration, with a optical convolution front-end in combination with an electronic implementation of the fully-connected layers \cite{chang2018hybrid,mengu2019analysis,colburn2019optical}. 

Programmable integrated photonics has made significant advances, potentially eliminating the need for such a hybrid system \cite{harris2018linear}. As compared to bulk optics, integrated photonics is a scalable solution in terms of alignment stability and total network size. Additionally, rapid advances in performance of integrated silicon photonics devices could enable such photonic CNNs to be faster and more energy efficient than electronic implementations \cite{atabaki2018integrating}. In fact, a recent proposal implements a photonic CNN by employing the patching technique to vectorize the input and kernel matrices \cite{bagherian2018chip}. However, the proposed architecture requires integrating long on-chip delay lines which requires overcoming some severe engineering challenges. In this work, we propose to use a star coupler (i.e. integrated diffractive optics) \cite{goodman2005introduction}, which implements the discrete Fourier transform (DFT), to perform the convolution operation. Combined with phase and amplitude masks, we construct an integrated photonics CNN. We present the performance of the photonic CNN on various datasets, study the performance degradation with imperfections and also provide discussion on scalability and possible future directions. 

\section{Photonic CNN}

A typical CNN architecture consists of convolution layers, pooling layers, activation layers, a fully-connected layer and finally an output layer. In this section, we first show the details of the star coupler DFT. Subsequently, we describe how to physically implement the convolution, pooling and activation layers using photonic components. Lastly, we present a variety of photonic CNN architectures together with prediction results on standard machine learning datasets. 

\subsection{DFT using star couplers}

\begin{figure}[tb]
	\centering
	\fbox{\includegraphics[trim={100 140 270 20},clip,width=\linewidth]{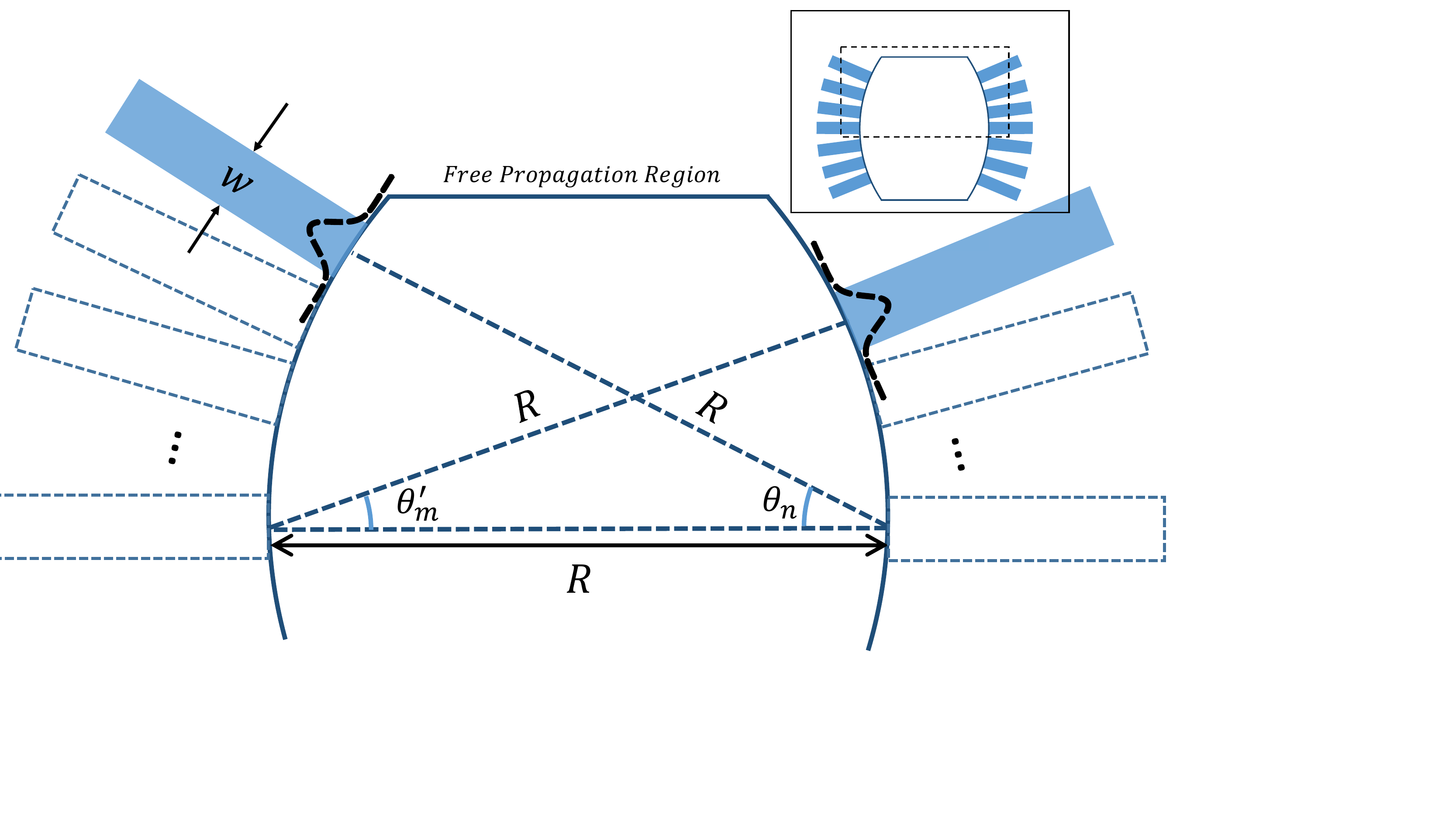}}
	\caption{Schematic of $N \times M$ star coupler. $R$ is the radius of the confocal circles that make up the free-space propagation region. $\theta_n$ is the angle of the $n$th input waveguide, $\theta_m'$ is the angle of the $m$th output waveguide. $w$ is the waveguide mode width parameter.}
	\label{fig:1}
\end{figure}

The unitary DFT operation $\mathcal{F}$ and inverse $\mathcal{F}^{-1}$ for a discrete signal $x[n]$ of length $N$ can be defined,

\begin{equation}
X[m] = \frac{1}{\sqrt{N}}\sum_{n=n_0}^{n_0+N-1} x[n] e^{-i2\pi\frac{nm}{N}}
\label{eq:1}
\end{equation}
\begin{equation}
x[n] = \frac{1}{\sqrt{N}}\sum_{m=m_0}^{m_0+N-1} X[m] e^{+i2\pi\frac{nm}{N}}
\label{eq:2}
\end{equation}

where $n_0$ and $m_0$ equals $-\frac{N}{2}$ if $N$ is even and $-\frac{N-1}{2}$ if $N$ is odd. Here, we use a star coupler to implement the DFT, as in \cite{takiguchi2011optical,cincotti2012else}.

A star coupler is a $N \times M$ device, with $N$ single-mode input waveguides and $M$ single-mode output waveguides connected by a ``free-space" propagation region like a slab waveguide \cite{dragone1989efficient}. The input and output waveguides are arranged along the circumference of two confocal circles of radius $R$. Under scalar diffraction theory and using the paraxial approximation, the coupling between an input waveguide at angle $\theta_n$ to an output waveguide at an angle $\theta'_m$ is \cite{klekamp2003calculation}

\begin{equation}
\kappa(\theta_n,\theta'_m) = U(\theta'_m) \int \Phi(\theta'-\theta'_m) e^{-i\tilde{k}R(\theta'-\theta'_m)\sin\theta_n} R d\theta'
\label{eq:3}
\end{equation}

where 

\begin{equation}
U(\theta'_m) = \frac{e^{i\tilde{k}R}}{\sqrt{i\tilde{\lambda}R}} \int \Phi(\theta-\theta_n) e^{-i\tilde{k}R\sin\theta\sin\theta_m'}R\cos\theta d\theta	
\label{eq:4}
\end{equation}

with $\tilde{k} = \frac{2\pi}{\lambda} n_s$ and $\tilde{\lambda} = \frac{\lambda}{n_s}$, $n_s$ being the slab effective index. $\Phi$ is the waveguide mode field, which we take to be a power normalized Gaussian, $\Phi(\theta) = \sqrt[4]{\frac{2}{\pi w^2}} e^{-(R\theta/w)^2}$ and $w$ is the width parameter of the waveguide. Since $w \ll R$, we can approximate $\theta' = \theta'_m$ and $\theta=\theta_n$ for the phase terms in Eq. \ref{eq:3} and \ref{eq:4}. 

Then, apart from a constant phase term,

\begin{equation}
\kappa(\theta_n,\theta'_m) \propto e^{-i\frac{2\pi}{N}\frac{NR}{\tilde{\lambda}}\sin\theta_n\sin\theta_m'}.
\label{eq:5}
\end{equation}

Comparing with Eq. \ref{eq:1}, assuming $N \ge M$, to get correspondence with DFT

\begin{eqnarray}
\theta_n & = & \sin^{-1} \left( n \sqrt{\frac{\tilde{\lambda}}{NR}}  \right) \label{eq:6a} \\
\theta'_m & = & \sin^{-1} \left( m \sqrt{\frac{\tilde{\lambda}}{NR}}  \right). \label{eq:6b}
\end{eqnarray}

Hence, by choosing the angular locations of the $n$th input and $m$th output waveguides by Eq. \ref{eq:6a} and \ref{eq:6b}, the star coupler can implement a DFT. 

\begin{figure}[tb]
	\centering
	\includegraphics[width=\linewidth]{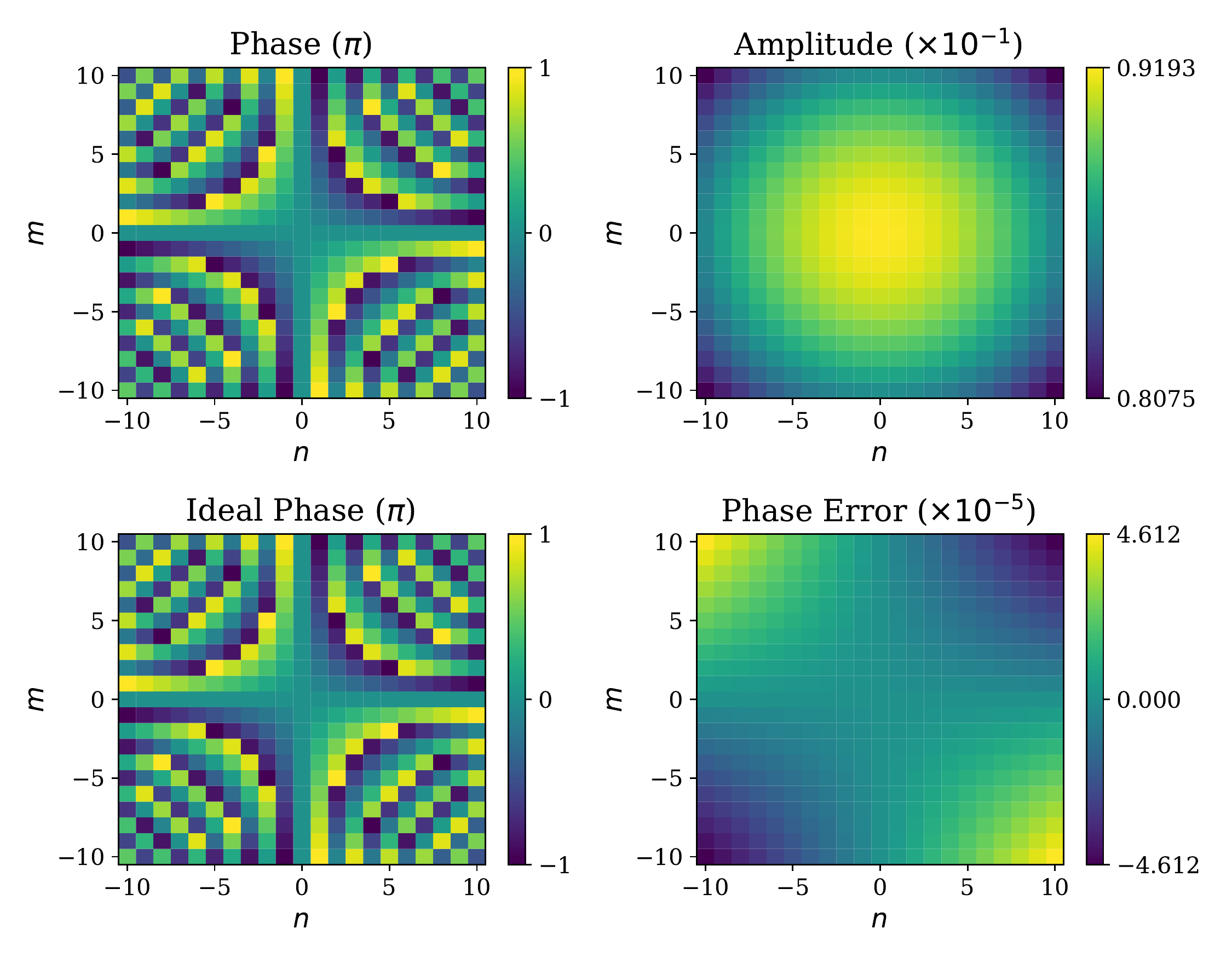}
	\caption{(Top) Phase and amplitude response of $21 \times 21$ star coupler DFT, calculated using Eq. \ref{eq:3}, with physical parameters: $\lambda=1550$ nm, $n_s = 2.85$, $w=500$ nm, $R=340.9~\mu$m. (Bottom) Ideal DFT phase response and relative phase error in radians of star coupler DFT.}
	\label{fig:2}
\end{figure}

In Fig. \ref{fig:2}, we plot the phase and amplitude response of a $21\times21$ star coupler DFT, $\mathcal{F}_{sc}$, calculated using Eq. \ref{eq:3}. Note that we have subtracted the constant phase term in the plot. The star coupler physical parameters are detailed in the figure captions and are chosen to ensure that the paraxial approximation is satisfied. Comparing with the ideal DFT, the greatest deviation in amplitude and phase response occurs at the waveguides furthest from the center. Adopting the fidelity measure as a distance metric \cite{fang2019design}, 

\begin{equation}
F = \left|\frac{\textrm{Tr}(||\mathcal{F}_{sc}||^\dagger\mathcal{F})}{N}\right|^2
\label{eq:7}
\end{equation}

gives $F=0.997$ for the star coupler described above, where $||\cdot||$ denotes division by the Frobenius norm. Another useful summary metric is the overall transmission

\begin{equation}
T = \frac{\textrm{Tr}(\mathcal{F}_{sc}^\dagger \mathcal{F}_{sc})}{N}
\label{eq:8}
\end{equation}

with $T = 0.162$ for the star coupler above. There is a trade-off between $T$ and $F$ and in a later section we will study this in more detail.  

We can compare the star coupler DFT with several other existing designs that implement the DFT with integrated photonics. Since the DFT is a unitary matrix, we can implement it directly using a mesh of Mach-Zehnder interferometers (MZIs) and phase shifters \cite{reck1994experimental,clements2016optimal}. Such a direct implementation of $N\times N$ DFT requires $\frac{N(N-1)}{2}$ MZIs, each consisting of twice that number of phase shifters and beam splitters. A more efficient design using the Cooley-Tukey FFT algorithm reduces the number of MZIs needed to $\frac{N\log_2(N)}{2}$ \cite{barak2007quantum,george2017towards}. However, the waveguide crossings needed grows as a triangular number $T_{\frac{N}{2}-1}$, which can introduce a significant insertion loss. Alternatively, 3D integration would be required to circumvent the need for crossings \cite{crespi2016suppression}. In comparison, the $N\times N$ star coupler DFT needs only a single "free-space" propagation region, which is a major reduction in complexity.

\subsection{Convolution, pooling and activation layers}

The convolution and pooling layers are linear and can be straightforwardly implemented optically. The convolution layer can be implemented optically in a ``4f" system, by using a cascade of two optical DFT operations with a phase and amplitude filter mask in between. Hence the convolution layer is defined $\mathcal{C}^N = \mathcal{F}_{sc}^{NN}\cdot A^{N} \cdot \mathcal{F}_{sc}^{NN}$, with the superscripts denoting the size of the matrices. Note that the second star coupler performs an inverse transformation back from the Fourier domain, except that the data is flipped vertically. Programmable phase and amplitude modulation can be applied to each individual waveguide using fast phase shifting mechanisms like thermo-optic and electro-optic effects \cite{orlandi2013tunable,watts2013adiabatic,timurdogan2017electric}. Alternatively, if a fast response is not essential, then reconfigurability using post-fabrication trimming or phase-change materials are attractive alternatives that do not need additional energy and control to maintain the state \cite{shen2011electric,topley2014planar,miller2018optical,feldmann2019all}. 

The pooling layer can be implemented as a low-pass filter \cite{rippel2015spectral}, by passing only the $M<N$ low-frequency components of the optical DFT, which correspond to the waveguides near to the central waveguide. Hence, the pooling layer is defined $\mathcal{P} = \mathcal{F}_{sc}^{MN}$. We may also combine convolution and pooling functions by transforming back from Fourier domain like so: $\mathcal{C}^{NM} = \mathcal{F}_{sc}^{MM}\cdot A^{M} \cdot \mathcal{F}_{sc}^{MN}$.

\begin{table}[htbp]
	\centering
	\caption{}
	\subcaption*{\bf Photonic CNN physical implementation}
	\begin{tabular}{ccc}
		\hline
		Network Layer & Operation & Optics \\
		\hline\hline
		Convolution  & DFT & $N \times N$ star coupler \\
		& Filter & Phase/Amp. mod. \\
		& DFT & $N \times N$ star coupler \\
		\hline 
		Pooling & DFT, Low-pass & $N \times M$ star coupler \\ 
		\hline
		Activation & modReLU &  O-E-O\\
		\hline
	\end{tabular}
	\label{tab:1}
\end{table}

Activation functions are nonlinear functions that allow the neural network to learn complex mappings between inputs and outputs \cite{hornik1989multilayer}. Nonlinear activation functions are regarded as one of the key reasons for the power of deep learning compared to classical machine learning methods. Adding nonlinearity into a photonic network substantially changes the functionality compared with previously demonstrated linear photonic circuits \cite{carolan2015universal,ribeiro2016demonstration,harris2018linear}. Several different kinds of optical nonlinearities have been proposed for implementation in optical neural networks, such as saturable absorption, optical bistability and two-photon absorption, to name a few \cite{nozaki2010sub,cheng2013plane,jiang2016analog}. However, all-optical nonlinearities are generally very weak and hence require high signal powers. As such, enhancing optical nonlinearities remains an area of active research. Recent advances in photonic integration have enabled demonstration of low-energy and high speed optical-electrical-optical or O-E-O devices, which act as pseudo-optical nonlinear devices \cite{nozaki2019femtofarad}. Such O-E-O devices are reconfigurable to show a variety of output responses, including an approximation of the widely used ReLU function \cite{tait2019silicon,williamson2019reprog}.

\begin{figure*}[tb]
	\centering
	\fbox{\includegraphics[trim={60 30 80 40},clip,width=0.7\textwidth]{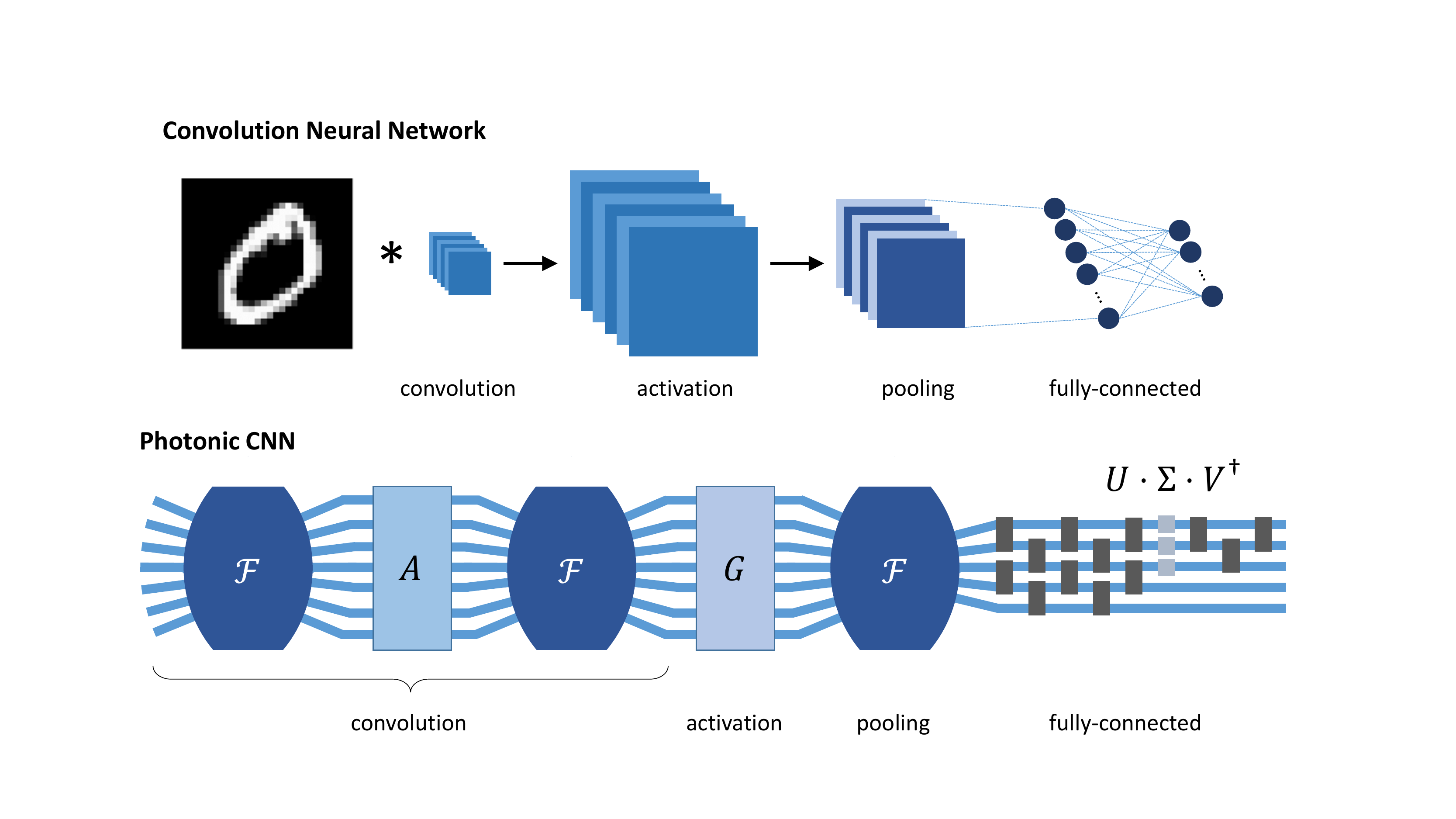}}
	\caption{Comparison between generic CNN architecture and corresponding photonic CNN implementation. The convolution is performed using a first star coupler DFT, then applying the filter mask in the Fourier domain and finally a second star coupler. Pooling is performed by passing only the low-frequency components through the output of the star coupler. Fully-connected layers are implemented as meshes of MZIs (dark gray boxes) and amplifiers/attenuators (light gray boxes).}
	\label{fig:3}
\end{figure*}

Figure \ref{fig:3} shows a comparison of a typical CNN architecture schematic and an equivalent photonic CNN implementation. We implemented the photonic CNN using TensorFlow \cite{abadi2016tensorflow} with the data encoded as the amplitudes of the complex field $u_0$ at the input waveguides. The filter mask $A$ is a diagonal matrix with complex-valued entries $a_n e^{i\phi_n}$, which are the trainable parameters for the convolution layers. This can be physically implemented as phase shifters and attenuators at each of the $n$ waveguides. For the pointwise activation function $G$, we formulate it as $\textrm{modReLU}(z_n,\{b_n,\varphi_n\}=0) = \mathrm{abs}(z_n)$, i.e. discarding the phase. See the Appendix for more details on modReLU. Hence, the complex field at the $(k+1)$th convolution layer is related to the $k$th layer as

\begin{equation}
u_{k+1}^M = G(\mathcal{C}_k^{MN} \cdot u_k^N) = G(\mathcal{F}_{k}^{MM}\cdot A_{k}^{M} \cdot \mathcal{F}_{k}^{MN} \cdot u_k^N)
\end{equation}

for a generic $N \times M$ convolution-pooling layer. Finally, we implemented the fully-connected layers in the same way as in Ref. \cite{shen2017deep}. Optically, it will be  

\begin{equation}
u_{k+1}^M = G(W_k^{MN}\cdot u_k^N) = G(U_k^{MM} \cdot \Sigma_k^M \cdot (V_k^{NN})^\dagger \cdot u_k^N)
\end{equation}

with $W$ as the weight matrix decomposed as $W=U\Sigma V^\dagger$ using singular value decomposition (SVD). More details of the implementation and network training procedures are found in the Appendix. 

\subsection{Results on MNIST dataset}

In Table \ref{tab:2}, we consider the performance of various different photonic ONNs on a standard machine learning task of digit recognition on handwritten images (MNIST) \cite{mnistdataset}. The 784 real-valued image pixels are fed into each of the 784 input waveguides using amplitude-only encoding. The PCNN-784 architecture is as follows: $\mathcal{C}^{784} \rightarrow \mathcal{C}^{392} \rightarrow \mathcal{C}^{196} \rightarrow W^{56} \rightarrow W^{10} $, where we omit the input size for clarity. We compare four different variations of PCNN-784 with modifications in \textit{only} the convolution layers: amplitude and phase modulation (i.e. $A_k^M = \textrm{diag}(a_me^{i\phi_{m}})$), amplitude-only modulation (i.e. $A_k^M = \textrm{diag}(a_m)$), phase-only modulation (i.e. $A_k^M = \textrm{diag}(e^{i\phi_{m}})$), and finally phase-only modulation with a linear activation function  (i.e. $\textrm{modReLU}(z_m,b_m=0)$). We find the best performance is obtained with amplitude and phase modulation, but only narrowly better than phase-only modulation. Amplitude-only modulation has the poorest performance. Below, we choose to focus on phase-only modulation since it achieves very good accuracy with less parameters, and it is also well adopted in existing literature \cite{chang2018hybrid,yan2019fourier,horner1984phase}. 

The training (test) accuracy obtained is $99.2\% (97.9\%)$, which is below the state-of-the-art test accuracy of $\approx99.6\%$. However, we expect the result to improve with advanced techniques like data augmentation, learning rate scheduling etc. We also tested the performance of the phase-only PCNN-784 on a more challenging dataset (F-MNIST) \cite{xiao2017fashion} and obtained an accuracy of $91.0\% (88.6\%)$ which is comparable to generic CNNs.

\begin{table}[htbp]
	\centering
	\caption{}
	\subcaption*{\bf Comparison of different ONN architectures}
	\begin{tabular}{lccc}
		\hline
		ONN & Param. & Data & Acc. (\%) \\
		\hline\hline
		PCNN-784 (Amp. \& Phase) & 14280 & MNIST & 99.6 (98.2) \\
		\texttt{"} (Amp.) & 12908 & MNIST & 97.4 (95.3)  \\
		\texttt{"} (Phase) & 12908 & MNIST & 99.2 (97.9)  \\
		\texttt{"} (Phase, Linear) & \texttt{"} & MNIST &  99.0 (97.6)\\
		\texttt{"} (Phase) & \texttt{"} & F-MNIST & 91.0 (88.6)\\
		\hline
		MLP-784 \cite{shen2017deep} & 12704 & MNIST & 95.9 (93.8)\\
		\hline
		D2NN-16 \cite{yan2019fourier} & 12544 & MNIST & 96.6 (95.6)\\
		\hline
		PCNN-256-32 & 4800 & MNIST & 97.7 (96.6)\\
		PCNN-256-16 & 2592 & MNIST & 96.1 (95.6)\\
		PCNN-112-32 & 2280 & MNIST & 93.9 (93.6)\\
		PCNN-112-16 & 1224 & MNIST & 91.2 (91.1)\\
		\hline
	\end{tabular}
	\label{tab:2}
\end{table}

To test the potential advantage of the PCNN architecture as compared to existing ONN architectures, we also simulate the performance of two other kinds of networks: MLP-784 is a fully-connected network ($W^{16} \rightarrow W^{10}$) of the kind found in Ref. \cite{shen2017deep}; while D2NN-16 is a stack of 17 star couplers ($\mathcal{F}_{sc}^{784}$) sandwiching 16 activation and phase layers, similar to Ref. \cite{yan2019fourier}. We intentionally chose the network architectures to have roughly the same number of trainable parameters for comparison. The results (Table \ref{tab:2}) show that the PCNN, which has a combination of convolution and fully-connected layers, indeed has an additional benefit in this classification task. 

Since the bulk of the trainable parameters in the PCNN come from the fully-connected layers, whereas the parameters in the convolution layers only grow linearly with the size of the input, we studied the trade-off between performance and complexity by being more aggressive with the pooling layers. The PCNN-x-y architecture is: $\mathcal{C}^{x} \rightarrow \mathcal{C}^{x/2} \rightarrow W^{y} \rightarrow W^{10}$, with phase-only modulation and linear activation in the convolution layers. We have introduced a pooling operation in the first layer to extract the low frequency components. A similar strategy of retaining only the low frequency features was adopted in Ref. \cite{williamson2019reprog,pai2019parallel}, in which the authors reported achieving a high accuracy of $98.9\% (97.8\%)$. There is some degradation of accuracy but potentially a great reduction in size and complexity of the network. 

\section{Component imperfections}

Here, we study the effect on performance of two kinds of imperfections in the photonic CNN: first, the inherent imperfection of the star coupler DFT; second, imperfection in implementation of pre-trained ideal components parameters.

\subsection{Imperfect DFT implementation using star couplers}

\begin{figure}[tb]
	\centering
	\includegraphics[trim={40 0 40 30},clip,width=\linewidth]{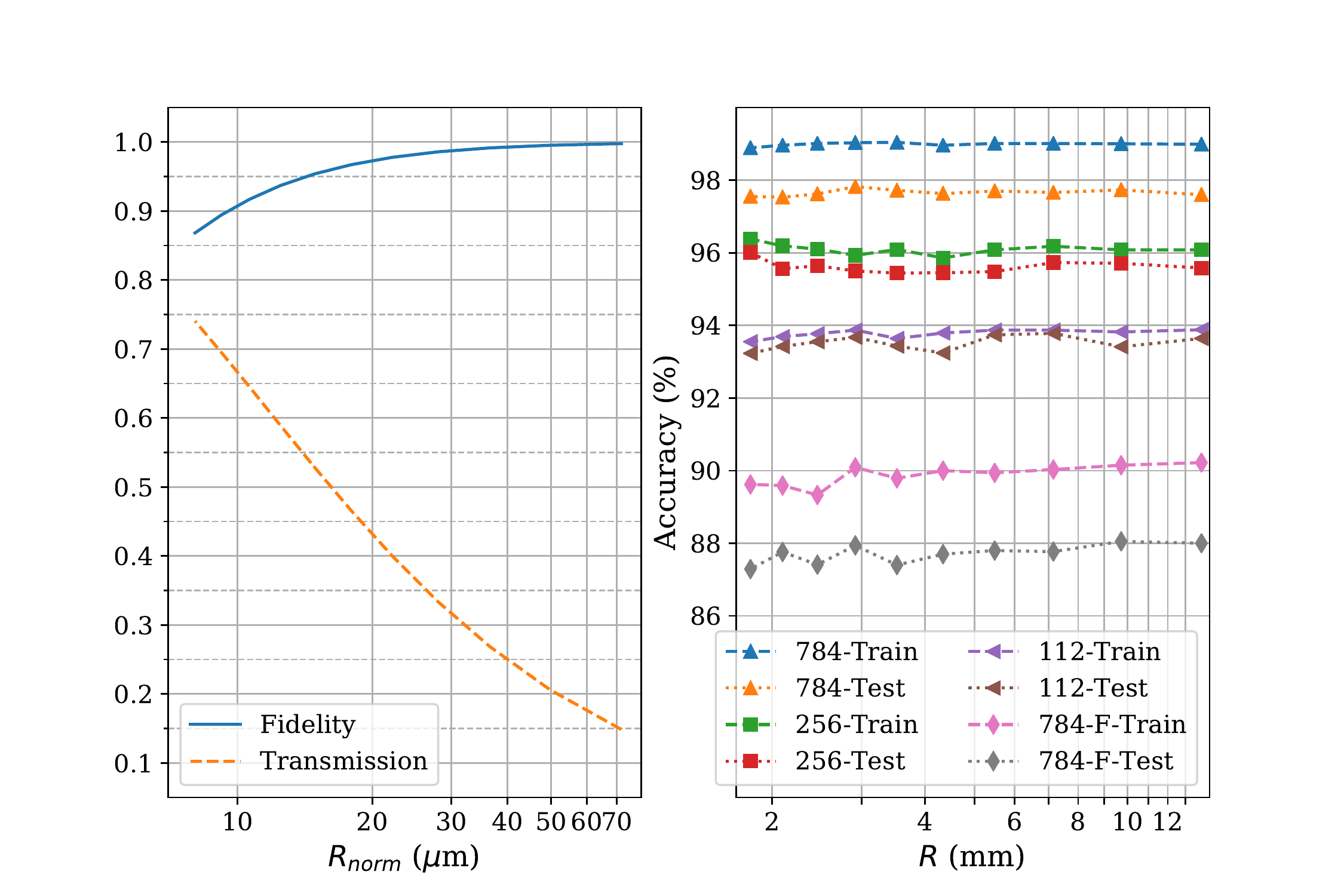}
	\caption{(Left) Trade-off between fidelity and transmission for star coupler DFT when varying normalized radius $R_{norm}$. (Right) Change in prediction accuracy versus input star coupler radius $R$ for different PCNN architectures, using the MNIST and F-MNIST data.}
	\label{fig:4}
\end{figure}

The radius $R$ of a $N \times N$ DFT star coupler is given by Eq. \ref{eq:6a} as,

\begin{equation}
\sqrt{R} = \frac{1}{|\sin \theta_{n_0}|} \sqrt{\frac{|n_{0}|^{2}}{N}\tilde{\lambda}}.
\label{eq:11}
\end{equation}

Since $N$ is determined by the size of the input, we have freedom to choose the angle of the waveguide furthest from the central line, $\theta_{n_0}$, as long as the paraxial approximation is satisfied. In Fig. \ref{fig:4}, we plot the fidelity $F$ and transmission $T$ of $784 \times 784$ star couplers with $5^{\circ}<\theta_{n_0}<15^{\circ}$, i.e. within paraxial limits. We defined a $N$ independent normalized radius $R_{norm} = \frac{\tilde{\lambda}}{|\sin \theta_{n_0}|^2}$, with $R \approx $ $\frac{N}{4} R_{norm}$. As can be seen, a very good fidelity is obtained with small $\theta_{n_0}$, but the transmission suffers and the star coupler radius is very large. By choosing a bigger $\theta_{n_0}$, there is much better transmission and nearly order of magnitude reduction in radius, with some sacrifice of fidelity. The reason for this trend is the following: choosing a small $\theta_{n_0}$ (large radius) concentrates the receiving waveguides near the axis and thus some of the optical power further away from the axis is not captured. On the other hand, choosing a large $\theta_{n_0}$ (small radius) angularly spreads out the receiving waveguides which are better able to cover the Gaussian shaped far-field envelope of the emitting waveguides. 

To study the effect of the reduced fidelity, we simulated different PCNN with varying $\theta_{n_0}$. In Fig. \ref{fig:4}, we plot the accuracy of these PCNN when trained on the MNIST and F-MNIST image recognition tasks. We observed almost no degradation of accuracy with increasing $\theta_{n_0}$ in the MNIST task and a small reduction of accuracy of about $0.6\%$ in the F-MNIST task. This indicates that the reduced fidelity can be compensated by the network training and is an advantageous trade-off for the gains in transmission and reduction in star coupler footprint.

\subsection{Non-idealities and fabrication imperfections}

\begin{figure}[tb]
	\centering
	\includegraphics[trim={0 0 0 0},clip,width=\linewidth]{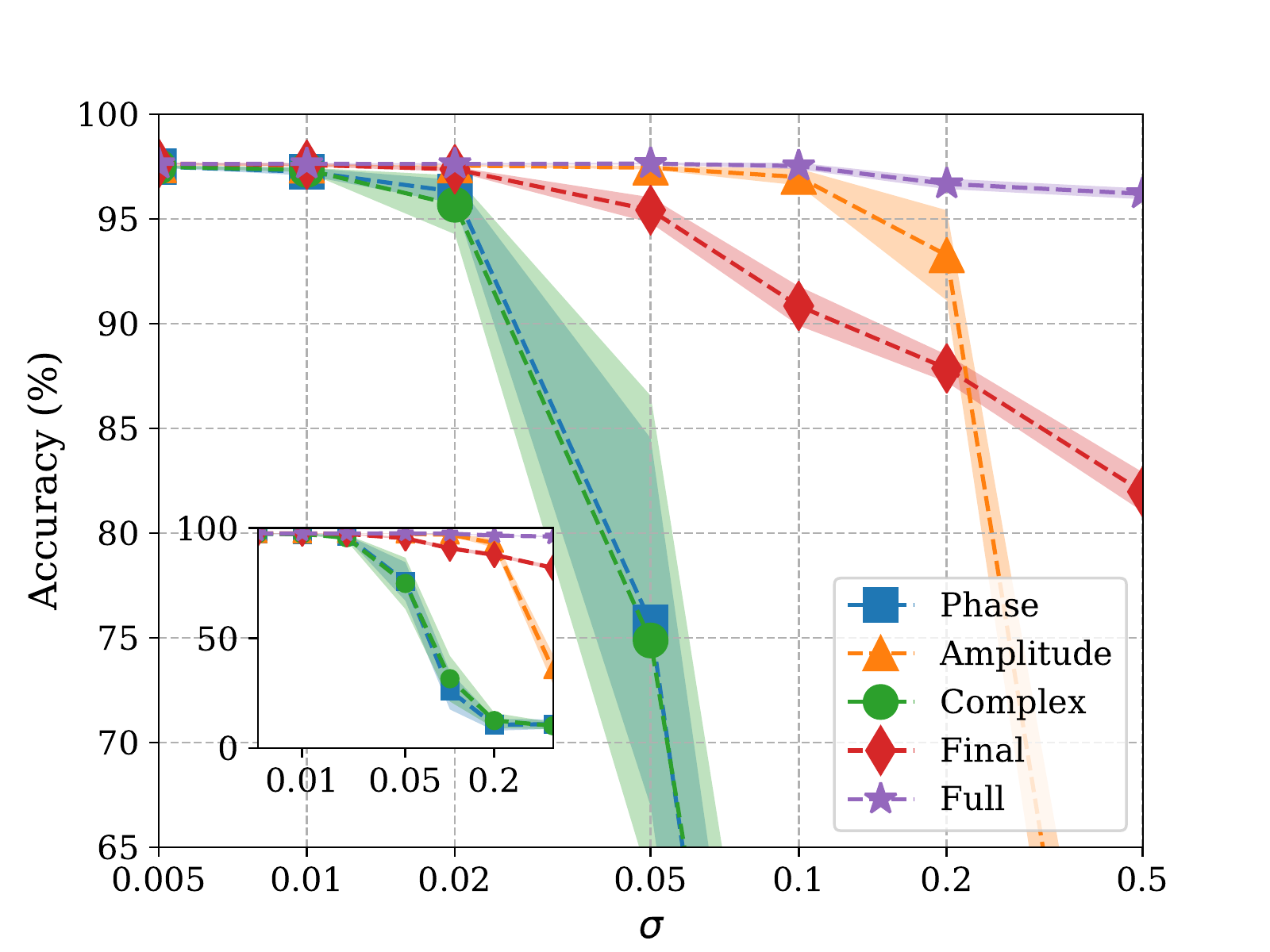}
	\caption{Degradation in prediction accuracy with increasing noise. Additional phase noise has standard deviation of $2\pi\sigma$, while amplitude noise is $a_\delta=1-|\delta|$ with $\delta$ having standard deviation of $\sigma$. ``Final" and ``Full" indicates restoration of accuracy after re-training the weights of the output layer and the full network when phase noise is added. (Inset) Zoom-out of plot, showing the complete randomization of the network when noise is large.}
	\label{fig:5}
\end{figure}

The current standard procedure to implement a optical neural network begins with training performed using a software simulation model of the system, followed by translation of the trained parameters to the optical device parameters and finally fabrication of the device. This method relies heavily on the accuracy of the mapping process from trained parameters to physical parameters. This includes, for example, non-idealities introduced by thermal cross-talk and the finite precision of electronic control circuits \cite{shen2017deep}. In addition, there will inevitably be fabrication imperfections of the photonic components (star couplers, waveguides, phase shifters, beam-splitters etc.) which break the correspondence between the trained software model and the hardware implementation \cite{flamini2017benchmarking,fang2019design}. Such additional uncertainty introduced by physical implementation becomes non-negligible especially when scaling to a large number of components, hence it is important to evaluate its effects on the photonic CNN performance. 

To study the effects of such imperfections, we evaluated the degradation of the MNIST classification accuracy of a pre-trained PCNN-784 (phase, linear) by introducing both amplitude and phase gaussian noise to the star coupler matrices $\mathcal{F}_{k}$ and the complex-valued filter masks $A_k$. We avoided adding noise to the fully-connected layers as this has been studied in previous literature along with several methods to ameliorate its effects being suggested \cite{flamini2017benchmarking,fang2019design}. The added phase noise $\Delta \phi$ is zero mean normally distributed with width $2\pi\sigma$, whereas the amplitude noise is modeled as an additional loss i.e. $a_{\delta} = 1 - |\delta|$, where $\delta$ is zero mean normally distributed with width $\sigma$. Hence, each element of the matrices $\mathcal{F}_k$ and $A_k$ is multiplied by a random complex factor $a_{\delta} \cdot e^{i\Delta\phi}$. 

Previous studies have considered phase noise of width up to $0.02~\textrm{rad}$, which is justified for low index contrast platforms \cite{flamini2017benchmarking,fang2019design}. However, for high index contrast platforms like silicon-on-insulator, the phase errors resulting from imperfections could be up to 2 orders of magnitude greater \cite{gehl17active} and hence we considered much larger phase errors. 

Figure \ref{fig:5} shows a comparison of the resulting degradation when adding purely phase noise, purely amplitude noise and complex-valued noise. We plot the mean and standard deviation of the accuracies, with each point in the plot consisting of 20 random instances. We can see from the results that the performance begins to degrade when $\sigma>0.02$. Additional phase noise is especially detrimental to the PCNN accuracy, as expected for such a coherent optical system, and it is imperative that mitigation strategies are in place for high index contrast platform like silicon photonics \cite{yang15phase,gehl17active}. Several such proposals and demonstrations of post-fabrication reconfiguration and optimization of programmable coherent optical meshes can be found in the literature \cite{annoni2017unscrambling,hughes2018training,shen2017deep,pai2019parallel}. As hoped, re-training PCNN-784 with added phase noise restores the network accuracy (see Fig. \ref{fig:5}), except in the most noisy configurations (see Appendix for details on training). Drawing inspiration from substantial previous work on randomly weighted networks in the literature \cite{schmidt1992feed,saxe2011random,huang2015extreme,saade2016random}, we attempted to restore the performance of noisy PCNN-784 by training only the weights of the output layer. Although the restored accuracy is reduced from the ideal case, we see a substantial improvement from the noisy state, which suggests that full reconfigurability may not be necessary for photonic CNNs to function despite the presence of a large amount of noise.

\section{Physical footprint}

Recently, progress has been made in demonstrations of silicon photonic integrated circuits with a large number ($\sim$1000s) of reconfigurable components \cite{sun2013large,han2015large,perez2017multipurpose,shen2017deep,harris2018linear,carolan2015universal,annoni2017unscrambling} and it should be possible to implement photonic neural networks with a comparable number of trainable parameters. As the complexity of the ONN increases, the physical footprint of photonics components will be an important limiting factor on scalability. As mentioned previously, the star coupler DFT potentially provides substantial reduction in footprint required. For concreteness, let us consider a $256\times 256$ DFT. A typical MZI will have physical size of $\sim 60 ~\mu \textrm{m} \times 100 ~\mu \textrm{m}$ \cite{williamson2019reprog}. Implementing the DFT using the Cooley-Tukey FFT algorithm requires 1024 MZIs, which will have a footprint of $6.14 ~\textrm{mm}^2$. In comparison, for a star coupler DFT of $R_{norm} = 10~\mu\textrm{m}$, the footprint would be $\sim0.6~\textrm{mm}\times0.3~\textrm{mm}$, which is 34 times smaller. This considerable reduction in footprint makes deep CNNs feasible to be implemented using photonic integrated circuits. 

%State-of-the-art CNNs can have millions of nodes and complicated interconnected architectures \cite{lecun2015deep,szegedy2017inception,ronneberger2015u}. Although there is a parallel effort towards simplifying networks for low-power applications \cite{hinton2015distilling,kim2015compression}, the photonic CNNs we have presented are still comparatively much less complex and hence further reduction in component footprint will enable implementation of more complicated CNN architectures. In particular, recent advances suggest there may be avenues to further shrink the physical footprint of the optical DFT through subwavelength structures or plasmonics \cite{fan2017integrated,wang2019chip}. Inverse design methods could offer even more drastic footprint reduction \cite{hughes2018adjoint,khoram2019nanophotonic}, but the potential trade-offs in loss and critical dimensions should be carefully considered. Other potentially advantageous tweaks include complex valued input encodings and fully unitary networks \cite{jing2017tunable,fang2019design}. 

\section{Conclusion}

In conclusion, we have proposed and simulated a scalable architecture for photonic convolutional neural networks using the Fourier transform property of star couplers. We described in detail methods to use photonic components to implement various layers of a generic CNN, including convolution and pooling. We compared our proposed architecture to existing designs and found a boost in performance as well as a significant reduction in complexity and footprint. We also considered effects of component imperfections and noted that the photonic CNN is robust to small amounts of noise and in the case of very noisy networks, required minimal re-training to restore network accuracy. 

Real implementations of photonic CNNs still requires important engineering work in practical areas such as latency, energy consumption etc. \cite{williamson2019reprog}. Fortunately, interest in the successes of deep learning has spurred significant efforts towards the realization of photonic neural networks, as evidenced by the numerous publications in recent years. We expect our proposed architecture to enable the implementation of scalable deep CNN on integrated photonic platforms and further the efforts towards the goal of fully optical neuromorphic computing platforms. 

%\cite{george2019neuromorphic,passalis2019training,zuo2019all}

\appendix
\label{sec:Appendix}

\subsection{Network training}

We implemented the photonic convolutional neural network (PCNN) using TensorFlow \cite{abadi2016tensorflow}. The PCNN consists of convolutional layers and fully-connected layers. Classification is done by taking the location of the maximum power at the output layer. 

\begin{figure}[tb]
	\centering
	\includegraphics[trim={0 0 0 0},clip,width=\linewidth]{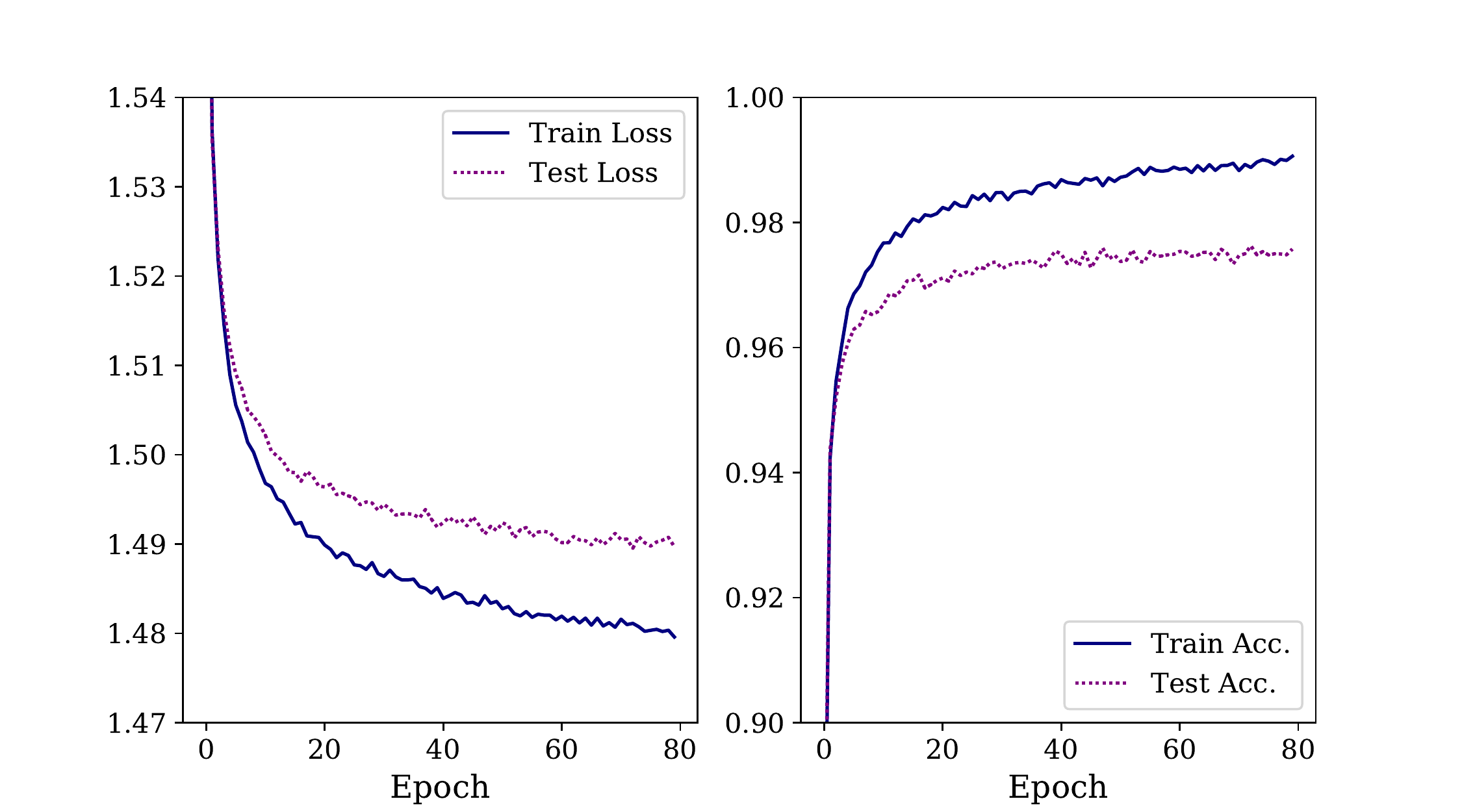}
	\caption{Loss and accuracy history over 80 epochs training of PCNN-784 (phase-only, linear) for MNIST training and test set.}
	\label{fig:s1}
\end{figure}

We used the stochastic gradient descent algorithm, Adam \cite{kingma2014adam}, for training. The loss function used for training is the cross-entropy with softmax function applied to the power at the output layer. The training batch size and the number of epochs was set to be 8 and 80, respectively. Figure \ref{fig:s1} shows an example of loss and accuracy training history. The MNIST and Fashion-MNIST datasets consist of $28\times28$ images split into a training set of 60,000 examples and a test set of 10,000 examples.

\subsection{Layer details}

The PCNN convolutional layers are made up of star couplers and filter masks. The $k$-th layer $N \times M$ star coupler coupling matrices $\mathcal{F}^{MN}_k$ are calculated using Eq. 3 and are fixed. Since $\mathcal{F}^{MN}_k$ has no trainable parameters, it is unchanged during network training. 

The $k$-th layer filter masks $A^{M}_k$ have complex-valued coefficients $a_{m} e^{i\phi_{m}}$. We define the filter mask coefficients $a_m$ and $\phi_m$ as \cite{mengu2019analysis}

\begin{eqnarray}
a_m & = & \frac{|\alpha_m|}{ \max\limits_{1 \leq m \leq M} |\alpha_m| }\\
\phi_m & = & 2\pi\theta_m
\end{eqnarray}

with $\alpha_m$ and $\theta_m$ being the trainable parameters. The normalization of $a_m$ ensures that it remains in the interval [0,1], whereas $\phi_m$ does not need normalization due to the periodicity of the phase $e^{i\phi_m}$. For phase-only modulation, we fix all $a_m = 1$.

For the nonlinear activation functions, we considered the modReLU function acting on the complex number $z_m = a_m e^{i\varphi_m}$ \cite{arjovsky2016unitary}. As an example, for the complex field vector $\mathbf{z}$ of the neural network layer, the activation on the $m$-th element

\begin{equation}
\textrm{modReLU}(z_m,b_m) = \textrm{ReLU}(a_m + b_m)\cdot e^{i\varphi_m}
\end{equation}

where the bias vector $\mathbf{b}$ is a trainable parameter. Table \ref{tab:s1} shows the accuracy results on the MNIST data for the phase-only PCNN-784 network when using different variations of modReLU. With $b_0$, we assume a single shared trained value for the bias vector $\mathbf{b}$. When $\{\mathbf{b},\pmb{\varphi}\}=0$, modReLU is equivalent to taking the $\textrm{abs}(z_m)$ as in the main text. Also, when $\mathbf{b}=0$, modReLU is just a linear activation. Table \ref{tab:s2} shows the effect of zero mean normally distributed bias noise of width $\Delta\mathbf{b}$. 

\begin{table}[htbp]
	\centering
	\caption{}
	\subcaption{\bf Comparison of different nonlinear activations}
	\begin{tabular}{llc}
		\hline
		Convolution & Fully-connected & Acc. (\%) \\
		\hline\hline
		modReLU($\mathbf{b}$) & modReLU($\mathbf{b}$) & 98.8 (96.7) \\
		modReLU($\mathbf{b}=0.01$) & modReLU($\mathbf{b}=0.01$) & 97.9 (96.6) \\
		modReLU($\mathbf{b}=b_0$) & modReLU($\mathbf{b}=b_0$) & 98.7 (96.6) \\
		modReLU($\mathbf{b}=b_0$) & modReLU($\{\mathbf{b},\pmb{\varphi}\}=0$) & 99.3 (97.7) \\
		modReLU($\mathbf{b}=0$) & modReLU($\mathbf{b}=0$) & 92.6 (92.4) \\
		\hline
	\end{tabular}
	\label{tab:s1}
	
	\bigskip
	\centering
	\subcaption{\bf Effect of bias noise $\Delta$b on accuracy of modReLU($\mathbf{b}$)}
	\begin{tabular}{llc}
		\hline
		$\Delta$b & Train Acc. (\%) & Test Acc. (\%) \\
		\hline
		\hline
		0.005 &  98.3$\pm$0.1 &  96.6$\pm$0.1 \\
		0.01 &  97.1$\pm$0.2 &  95.7$\pm$0.2\\
		0.02 &  92.7$\pm$0.9 &  91.9$\pm$0.9\\
		0.05 & 63.7$\pm$3.4 &  63.8$\pm$3.5\\
		0.1 & 36.6$\pm$3.1 &  37.2$\pm$3.0\\
		0.2 & 23.0$\pm$3.6 &  23.5$\pm$3.6\\
		0.5 & 16.9$\pm$3.7 &  17.3$\pm$4.0\\
		\hline
	\end{tabular}
	\label{tab:s2}
\end{table}

%\begin{figure*}[tb]
%	\centering
%	\fbox{\includegraphics[trim={0 0 0 0},clip,width=1.00\textwidth]{usvh.pdf}}
%	\caption{$196 \times 56$ weight matrix from a trained PCNN-784 (phase-only) and its SVD.}
%	\label{fig:s2}
%\end{figure*}

The $k$-th fully-connected layers of the PCNN are described by a real-valued weight matrix $W_k^{MN}$, which can be decomposed by SVD into a product of two unitary matrices and a non-negative real diagonal matrix, i.e. $W = U\Sigma V^\dagger$. For the purposes of training the network, we take the tunable parameters as the real-valued elements of $W$, with the knowledge that it can be implemented optically through the decomposition \cite{shen2017deep}. In general, attenuators can be used to implement a scaled matrix $\Sigma'=\frac{\Sigma}{\beta}$, such that the singular values are $\leq1$. In that case, a global optical amplification $\beta$ is needed and the weight matrix is $W = \beta \cdot U\Sigma ' V^\dagger$ \cite{fang2019design}. In the main text, the choice of activation function is multiplicative, i.e. $\textrm{abs}(\beta z) = \beta\cdot\textrm{abs}(z)$. Hence, using a scaled weight matrix $W' = U\Sigma ' V^\dagger$ would give the same prediction result as using $W$.

%Figure \ref{fig:s2} shows an example of a $196 \times 56$ weight matrix from a trained PCNN-784 (phase-only) and its SVD. 

\subsection{Re-training noisy networks}

For the re-training of noisy networks, the training batch size and the number of epochs was set to be 8 and 10, respectively. Gaussian noise was added to the star coupler matrices $\mathcal{F}_k$ and the complex-valued filter masks $A_k$. During re-training, only the $A_k$ and $W_k$ matrices are trainable, while the $\mathcal{F}_k$ are fixed in their noisy state. 

%Figure \ref{fig:s3} shows an example of PCNN-784 (phase-only) classification output and phase/weight matrices in three cases: the ideal case, after adding phase noise and after re-training. 

%\begin{figure*}[tb]
%	\centering
%	\fbox{\includegraphics[trim={0 0 0 0},clip,width=1.00\textwidth]{figure_s30.pdf}}
%	\fbox{\includegraphics[trim={0 0 0 0},clip,width=1.00\textwidth]{figure_s31.pdf}}
%	\caption{(Top) Classification result for a sample input in the noisy network ($\sigma=0.1$) and after re-training. Noisy network accuracy: 23.5\% (23.7\%), Re-trained network accuracy: 98.3\% (97.5\%) (Bottom) %Phase values of filter masks and weight values of fully-connected layers in PCNN-784 (phase). (Left) Ideal weights. (Middle) Weights with added noise. (Right) Weights after re-training.}
%	\label{fig:s3}
%\end{figure*}

%\section*{Acknowledgment}
%The authors acknowledge the usage of computational resources of the National Supercomputing Centre, Singapore %(https://www.nscc.sg) for this work.

\bibliographystyle{IEEEtran}
\bibliography{IEEEabrv,ref}

\end{document}